%% file: main.tex
\documentclass[journal]{IEEEtran}
\usepackage{graphicx}
\usepackage{color}
\usepackage{cite}
\usepackage{amsmath,amssymb}
\usepackage{threeparttable}

\usepackage{pgfplots}
\pgfplotsset{compat=1.9}
\usepackage{tabularx,booktabs}
\usepackage{wasysym}
\usepackage[bookmarks=false]{hyperref} 
\usepackage{float}
\newcolumntype{C}{>{\centering\arraybackslash}X} 
\usepackage{siunitx}
\usepackage{caption}

\usepackage{tikz}
\usepackage{tikzscale}

\usepackage{dblfloatfix}
\usepackage{epstopdf}
\AppendGraphicsExtensions{.tif}

\makeatletter
\renewcommand*\env@matrix[1][\arraystretch]{%
  \edef\arraystretch{#1}%
  \hskip -\arraycolsep
  \let\@ifnextchar\new@ifnextchar
  \array{*\c@MaxMatrixCols c}}
\makeatother

\usepackage{color}

\usepackage{makecell}

\begin{document}

\title{A Broadband Conversion Loss Measurement Technique for Terahertz Harmonic Mixers}

\author{Divya~Jayasankar,~\IEEEmembership{Graduate Student~Member,~IEEE,}
    Theodore Reck,~\IEEEmembership{Senior Member,~IEEE,}
    Steven Durant, 
    Jan~Stake,~\IEEEmembership{Senior Member,~IEEE,}
    and Jeffrey~Hesler,~\IEEEmembership{Fellow,~IEEE}

\thanks{Manuscript received 24 October 2023; revised 4 January 2024; accepted Jan xxxx. Date of publication Jan xxxx; date of current version Feb xxxx. This research was carried out in the GigaHertz-ChaseOn bridge center at the Chalmers University of Technology in a project financed by VINNOVA, Chalmers University of Technology, AAC Clydespace, Low Noise Factory and Virginia Diodes, Inc. (VDI). Miss. Jayasankar's PhD project is supported by the Swedish Foundation for Strategic Research (SSF) project number FID17-0040 and the Swedish Innovation Agency (Vinnova) project 2022-02967. Her research visit to Charlottesville was partly funded by the Roger Pollard Student Fellowship from the Automatic RF Techniques Group (ARFTG). \textit{(Corresponding author: Divya Jayasankar, divyaj@chalmers.se).} }
\thanks{D. Jayasankar and J. Stake are with the Terahertz and Millimetre Wave Laboratory, Department of Microtechnology and Nanoscience (MC2), Chalmers University of Technology, SE-412 96 Gothenburg, Sweden. D. Jayasankar is also with Research Institutes of Sweden (RISE), SE-504 62 Borås, Sweden.}
\thanks{T. Reck, S. Durant, and J. Hesler are with the Virginia Diodes Inc. (VDI), Charlottesville, VA 22902, USA.}
\thanks{Color versions of one or more figures in this article are available online at http://ieeexplore.ieee.org. Digital Object Identifier 10.1109/TTHZ.2024.xx}}

\markboth{To be submitted to IEEE TRANSACTIONS ON TERAHERTZ SCIENCE AND TECHNOLOGY, 2023}%
{Divya \MakeLowercase{\textit{et al.}}}

\markboth{Submitted to IEEE TRANSACTIONS ON TERAHERTZ SCIENCE AND TECHNOLOGY, 2023}%
{Divya \MakeLowercase{\textit{et al.}}}

\maketitle

\begin{abstract}

 This letter presents an experimental characterization technique for assessing the performance of terahertz harmonic mixers across a wide frequency range. The total signal transfer loss of three mixers was measured in both up- and down-conversion configurations, and the conversion loss was determined through the solution of a linear system of equations. \textcolor{black}{The proposed method uses LO signals with a frequency offset to ensure single sideband measurements, thereby eliminating the need for image-reject filters.}  The three-mixer method was verified by measurements of millimeter-wave mixers, which matched the traditional characterization method using a calibrated source and power meter. Given this successful millimeter-wave demonstration, we characterized three WM-86 Schottky diode ×4-harmonic mixers from 2.2 to 3 THz. This technique presents a notable advantage for conducting broadband mixer characterizations, particularly in the terahertz frequency regime which lacks tunable, wide-band sources.


\end{abstract}


\IEEEpeerreviewmaketitle

\input{Chapter/Introduction}

\input{Chapter/Methodology}

\bibliographystyle{IEEEtran}
\bibliography{IEEEfull,bibl}

\end{document}

%% file: Chapter/Introduction.tex
\section{Introduction}

\IEEEPARstart{F}{requency} translation devices, commonly known as mixers, play a pivotal role in radio systems and heterodyne instrumentation \cite{Herold1942}. Notably, harmonic mixers have recently gained significant prominence and have become integral components in various test and measurement instruments, including signal and spectrum analyzers, frequency counters, and frequency stabilization systems \cite{hayton2013, Curwen2023}.

Since the advent of crystal diode rectifiers for frequency conversion in the early 1900s, extensive research has been conducted to assess mixer performance accurately. Traditionally, mixer conversion loss has been evaluated through techniques such as the Y-factor power-meter method \cite{maas}, variable attenuator-method \cite{Trambarulo}, and scalar and vector mixer calibration approach \cite{Dunsmore,Alireza2013}. However, mixers utilizing higher-order harmonics exhibit high conversion loss \cite{bulcha2016, divya2021, Reck2023}, rendering the Y-factor method impractical. Furthermore, the absence of calibrated power sources at terahertz (THz) frequencies also imposes limitations on mixer characterization using the aforementioned conventional methods. Although the emergence of far-infrared optical sources like quantum-cascade lasers (QCLs) has accelerated progress in THz mixers, the lack of tunable and wide-band sources remains a significant constraint. This limitation underscores the critical need for an alternative measurement technique to characterize THz mixers across a broad frequency range.

\begin{figure}[t!]
    \centering
    \includegraphics[width=0.995\linewidth]{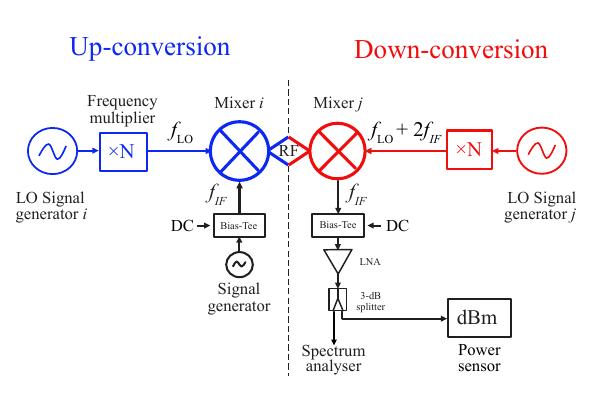}
    \caption{\textcolor{black}{Mixer characterization method. Schematic showing the up-down conversion setup.}}
    \label{schematic}
\end{figure}

In this letter, we adopt the  \lq three-antenna measurement method\rq~\cite{Newell73} proposed by Beatty in 1967 \cite{Beatty67} for characterizing THz harmonic mixers. We provide a detailed explanation of the theoretical underpinnings governing the conversion loss extraction, assuming reciprocal conditions in section II. The latter part of this section delineates the experimental arrangement for single sideband (SSB) measurements, which features WM-86\footnote{RF waveguide name designation. 'W' stands for waveguide, 'M' for metric, and the number is the waveguide width in \textcolor{black}{micrometers} \cite{waveguide}.} $\times$4-harmonic mixers with integrated diagonal horns arranged head-to-head and are driven by solid-state LO sources. Section III presents broadband conversion loss measurements of WM-86 $\times$4-harmonic mixer spanning from 2.2 to 3~THz. Finally, we validate the effectiveness of this method by comparing it with the conventional measurement technique within the millimeter-wave frequency range.

%% file: Chapter/Methodology.tex
\section{Method}
\label{sectionII}
\noindent The conversion loss of a mixer can be found by measuring the total signal transfer loss, $L_{t} = L{_i^{up}} \cdot L{_j^{down}}$, through an up-conversion (IF to RF) in mixer $i$, which is directly down-converted in mixer $j$ (RF to IF) for a set of $N$ different mixers; see Fig.~\ref{schematic}. To ensure single-sideband (SSB) measurements, the mixer's image response is filtered out using two local oscillator signals (LO\textcolor{black}{$_i$} and LO\textcolor{black}{$_j$}) with a frequency difference twice the intermediate frequency ($2 \times f_{IF}$), as shown in \ref{schematic}. This approach entails utilizing the upper sideband as an RF signal for one of the mixers and the lower sideband for the other. The IF signal is kept constant, and the two synchronized LO signals are swept to perform a broadband conversion loss measurement. To minimize errors due to re-connections of the LO signal, we have utilized dedicated LO sources, each one paired with a specific mixer (DUT), see Fig.~\ref{schematic}. In this way, the mixer's pump (LO) and bias conditions, critical for the mixer's performance, are kept the same during different measurements.

Most terahertz waveguide mixes include an integrated horn antenna to reduce the RF waveguide loss and to avoid the effects of flange misalignment \cite{Li2014}. However, due to the complexity of the implementation of quasi-optical systems \cite{goldsmith} and the losses incurred by atmospheric attenuation at THz frequencies, we decided to employ a straightforward head-to-head alignment for the horns. This alignment strategy presents two drawbacks: it introduces a few decibels of coupling loss and generates standing waves within the RF chain. It is, therefore, not feasible to compensate for systematic errors due to standing waves between the RF ports. Additionally, the return loss at the IF ports is also neglected. Furthermore, 'magnitude' reciprocity is assumed for the mixer up and down conversion loss, ($|L{_i^{up}}|=|L{_i^{down}}|$), which is a valid approximation for a passive mixer \cite{Vanmoer2007}.

Using three mixers (X, Y, and Z), the signal transfer, $L_{t}$, is measured in both directions for each combination of mixer pairs, resulting in six measurements. In logarithmic scale (dB), the set of measurements results in a linear system of equations, $A L = L_{t}$, where $A$ is a coefficient matrix, $L$ is a vector with each mixer's conversion loss, and $L_{t}$ contains total measured signal loss. Since we restricted the number of re-connections, the mixer conditions are almost identical, except for one mixer tested under two LO conditions (LO$_i$ and LO{$_j$) for either lower or upper sideband mode. Hence, it is necessary to include an extra unknown, and, in this case, the overall equation system contains six measurements and four unknowns, resulting in a non-square matrix \ref{equation1}. 

A least-square approximation that minimizes the error, defined as ${|A L - L_{t}|}^2$, can be found by solving $A^{T}A L = A^{T}L_{t}$. If the upper and lower sideband conversion loss is assumed to be identical, three measurements are enough to provide an analytical solution \cite{Fujii2012, Tosaka2012}. However, the least square technique helps to suppress the effect of measurement uncertainties, and we found that the system with four unknowns resulted in a much lower ripple across the frequency band and, therefore, the primary method used in the article.

\begin{equation}
\begin{bmatrix}
\label{equation1}
\textcolor{black}{1}  & \textcolor{black}{0} & \textcolor{black}{1} & \textcolor{black}{0} \\
\textcolor{black}{1}  & \textcolor{black}{0} & \textcolor{black}{1} & \textcolor{black}{0} \\
\textcolor{black}{0}  & \textcolor{black}{1} & \textcolor{black}{0} & \textcolor{black}{1} \\
\textcolor{black}{0}  & \textcolor{black}{1} & \textcolor{black}{0} & \textcolor{black}{1} \\
\textcolor{black}{0}  & \textcolor{black}{0} & \textcolor{black}{1} & \textcolor{black}{1} \\
\textcolor{black}{0}  & \textcolor{black}{0} & \textcolor{black}{1} & \textcolor{black}{1} \\

\end{bmatrix}
\begin{bmatrix}[1.5]
\textcolor{black}{L{_X^{USB}}} \\ \textcolor{black}{L{_X^{LSB}}} \\ \textcolor{black}{L{_Y^{LSB}}} \\ \textcolor{black}{L{_Z^{USB}}}
\end{bmatrix}
=
\begin{bmatrix}[1.5]
\textcolor{black}{L{_t^{1}}} \\ \textcolor{black}{L{_t^{2}}} \\ \textcolor{black}{L{_t^{3}}} \\ \textcolor{black}{L{_t^{4}}} \\ \textcolor{black}{L{_t^{5}}} \\ \textcolor{black}{L{_t^{6}}}
\end{bmatrix}
\end{equation}

\subsection{Experimental setup}

The device-under-test (DUT) is a $\times4$-harmonic, single-ended, integrated GaAs Schottky diode mixer from VDI, based on the design in \cite{Reck2023}. The mixers were dc-biased using a constant current source. The mixer circuitry is assembled in an E-plane split block with an integrated WM-86 rectangular waveguide \textcolor{black}{and a standard diagonal horn of about 5-mm long, 0.56-mm $\times$ 0.56-mm aperture} \cite{Johansson_1992}. The two mixers are aligned head-to-head as shown in Fig.~\ref{setup}. 

\begin{figure}[hb]
    \centering
    \includegraphics[width=0.995\linewidth]{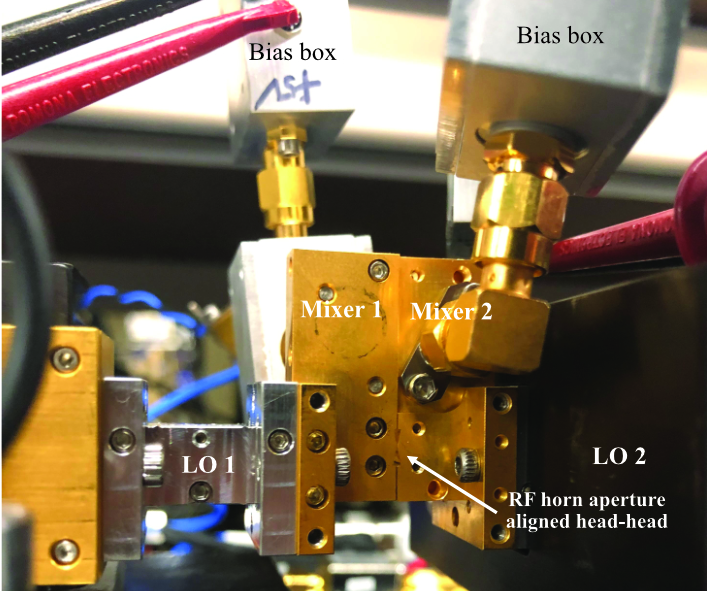}
    \caption{\textcolor{black}{Measurement setup. Photograph of the WM-86 $\times$4-harmonic mixers with integrated diagonal horn aligned head-to-head. The mixers are driven by WM-380 $\times$54 active multiplier chain (AMC) LO sources from VDI \cite{VDI}.}  
  }
    \label{setup}
\end{figure}

The WM-86 $\times$4-harmonic are pumped by WM-380 $\times$54 AMC source with available output power about 0.3-2~mW\footnote{The available output power of the three LO multiplier chains was measured using a VDI Erickson power meter PM5 and was corrected for the loss in the WR-10 taper.} in the frequency range 550--750 GHz \cite{VDI}. For up-conversion, 1-GHz IF signal generated by Keysight MXG analog signal generator N5183B with a power level of -20 dBm was fed to the mixer via a bias-tee. To pump the mixer, the signal from one of the sources in Keysight vector network analyzer PNA-X N5222B with direct digital synthesizer (DDS) was swept in frequency and multiplied by WM-380 $\times$54 AMC source to generate the LO signal from 550\ to 750 GHz. The RF signal generated by the up-conversion mixer was coupled to the down-conversion mixer, as shown in Fig.~\ref{setup}. The down-converted IF signal from the mixer was amplified by low-noise amplifiers (LNA) with 60-dB gain and was evaluated in a spectrum analyzer. The insertion loss of the cables, LNA, band-pass filter, and 3-dB splitter used in this measurements were characterized and de-embedded \cite{Penfield}. The signal generator, VNA, and spectrum analyzer were phase-locked to a common microwave reference signal.

\begin{figure}[ht!]
    \centering
    \input{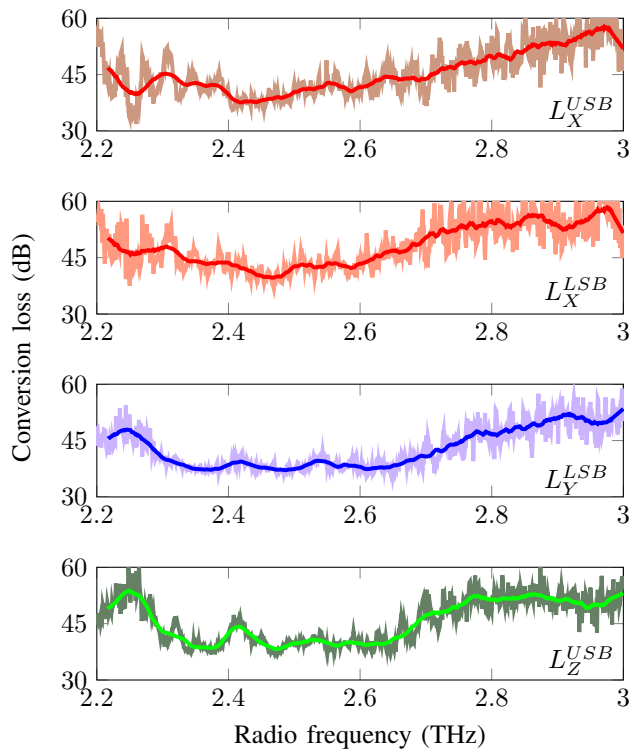}
    \caption{Conversion loss of three WM-86 $\times$4- harmonic mixers extracted using the least-square method plotted versus frequency. The shaded lines are the raw extracted data, and the colored lines are the smoothed data using the Savitzky-Golay finite impulse response (FIR) filter with a polynomial order of 1 and frame length 35 \cite{Savitzky1964}. Note: USB-Upper sideband, LSB-Lower sideband. }
    \label{LS}
\end{figure}

\section{Results}

\subsection{Sub-mm wave mixer measurements}
Based on the extraction procedure described in Section II, conversion loss of three WM-86 $\times$4-harmonic mixers were evaluated from 2.2 to 3 THz as shown in Fig.~\ref{LS}. Mixer 'X' utilized both upper and lower side-bands during the up- and down-conversion process, which resulted in four unknowns. Additionally, the observed ripples in the data are attributed to standing waves and parasitic modes in the \textcolor{black}{RF chain \cite{Morgan2013}, which is the main source of error for this specific measurement.} \textcolor{black}{This was verified by electromagnetic (EM) simulation of two horns aligned head-to-head, which resulted in an insertion loss ripple in the order of 5~dB}. Still, the smoothed data \cite{Savitzky1964} represents a reasonable estimate of the harmonic mixer conversion loss. It is also notable that as the conversion loss surpasses 50~dB, the noise levels increase, eventually approaching the signal-to-noise ratio (SNR) limit of the setup. Therefore, the insufficient link budget in this frequency range cannot allow any attenuator or free-space measurements \cite{Friis1946} to overcome the standing waves in the RF chain.

\subsection{Verification at mm-wave}

To verify the method outlined in section II, we conducted a characterization study in the millimeter-wave spectrum spanning from 50 to 75 GHz. Three WR-15\footnote{WR - Rectangular Waveguide, the number is the waveguide width in mils multiplied by 10. (1 mils = 1/1000 inch).} sub-harmonic ($\times$2) mixers \cite{VDI2} were characterized in the experimental configuration depicted in the inset of Fig.~\ref{het}. To mitigate the impact of standing waves, an isolator was introduced between the mixers. The resultant outcomes, obtained by solving the system of linear equations \cite{Fujii2012}, were compared with a single-mixer measurement performed with a calibrated source. The comparison, illustrated in Fig.~\ref{het}, demonstrates good agreement, with discrepancies within $\pm$~1~dB across the frequency band.

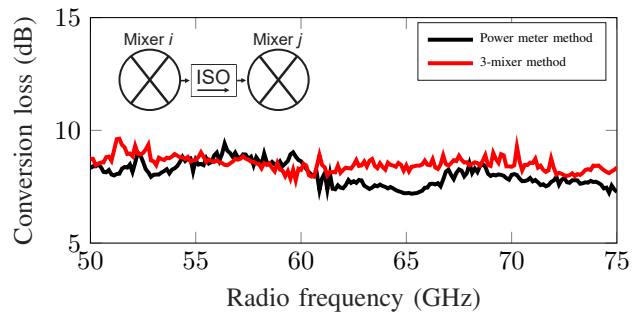
\begin{figure}[ht!]
    \centering
    \input{figures/heter}
    \caption{Comparison of analytically extracted conversion loss using the three-mixer method and conventional measurement of a single mixer with a calibrated source and power meter.}
    \label{het}
\end{figure}

\section{Conclusion}
We demonstrated the first experimental characterization of THz mixers from 2.2 to 3 THz. A key aspect of this technique is utilizing LO signals with a frequency offset to ensure SSB measurements, thereby eliminating the need for image-reject filters, which are challenging to implement at THz frequencies. We adopted a straightforward approach to simplify the alignment process, albeit at the cost of additional losses due to sub-optimal coupling and the occurrence of standing waves in the RF chain. A detailed uncertainty analysis of the proposed method requires a well-defined RF coupling and, preferably, IF/RF mismatch measurements, which were not feasible in this work. This technique proves to be highly effective for mixer characterization, particularly in the terahertz frequency region where there is a shortage of broadband, tunable sources.



%% file: figures/heter.tex
%
%

\definecolor{mycolor3}{rgb}{1,0,0}%
\definecolor{mycolor2}{rgb}{0,0,0}%
\begin{tikzpicture}

\begin{axis}[%
width=7cm,
height=3cm,
at={(0.817in,0.82in)},
scale only axis,
xmin=50,
xmax=75,
xlabel style={font=\color{white!15!black}},
xlabel={Radio frequency (GHz)},
ymin=5,
ymax=15,
ytick = {5,10,15},
ylabel style={font=\color{white!15!black}},
ylabel={Conversion loss (dB)},
axis background/.style={fill=white},
legend style={legend cell align=left, font = \tiny, align=left, draw=white!15!black}
]

\addplot [color=black, solid,line width=1.5pt]
  table[row sep=crcr]{%
50	8.3278154686178\\
50.125	8.407071413474\\
50.25	8.4057307369112\\
50.375	8.4931091788088\\
50.5	8.3694342383411\\
50.625	8.1562445756529\\
50.75	8.1368559464805\\
50.875	8.2410005165753\\
51	8.043422751637\\
51.125	7.9993869786102\\
51.25	8.0641310964846\\
51.375	8.0765372785653\\
51.5	8.3034118579028\\
51.625	8.3750785389762\\
51.75	8.3488100275192\\
51.875	8.4884060391539\\
52	8.1561055616907\\
52.125	8.566329875182\\
52.25	9.0189098212995\\
52.375	8.8946007344812\\
52.5	8.6031088521857\\
52.625	8.2775109088339\\
52.75	8.1359280442427\\
52.875	7.9391897220802\\
53	8.0307185007295\\
53.125	8.0399294557155\\
53.25	8.0409579694861\\
53.375	8.1283847809731\\
53.5	8.1939269959821\\
53.625	8.2761899168724\\
53.75	8.0902060430239\\
53.875	8.2590495578601\\
54	8.3148859524314\\
54.125	8.4207738348716\\
54.25	8.5985263386205\\
54.375	8.8257389488596\\
54.5	8.6399665994654\\
54.625	8.7488442085722\\
54.75	8.7620549849746\\
54.875	8.1819994390745\\
55	8.6885879742872\\
55.125	8.4828887933343\\
55.25	8.5087630546403\\
55.375	8.4548695318594\\
55.5	8.7576530853872\\
55.625	8.531622970086\\
55.75	8.5933305202732\\
55.875	8.8853721937251\\
56	9.0110491975283\\
56.125	8.6547681492263\\
56.25	8.955517909576\\
56.375	9.4143073558464\\
56.5	9.1035560752412\\
56.625	9.000834731881\\
56.75	8.8342783081682\\
56.875	9.013755723759\\
57	8.7965673342637\\
57.125	8.7758650532213\\
57.25	8.9359059354977\\
57.375	8.9261685132455\\
57.5	8.7373258471917\\
57.625	8.8277343343312\\
57.75	8.4529441499896\\
57.875	8.8088027905629\\
58	8.6434173623576\\
58.125	8.498125332191\\
58.25	8.9725134547747\\
58.375	8.5807049211704\\
58.5	8.4788174645848\\
58.625	8.6003537886556\\
58.75	8.4900986904364\\
58.875	8.4708714620916\\
59	8.6233093785133\\
59.125	8.4115538438606\\
59.25	8.5249599196282\\
59.375	8.6415363876602\\
59.5	8.9369183902086\\
59.625	9.0254888451079\\
59.75	8.8361413145072\\
59.875	8.6891177760085\\
60	8.6312165781275\\
60.125	8.4504487056785\\
60.25	8.2916948677397\\
60.375	8.1417637433252\\
60.5	8.0369685443673\\
60.625	7.9585710828474\\
60.75	8.0867072878859\\
60.875	7.3463301336437\\
61	7.8721042558744\\
61.125	8.098415474664\\
61.25	7.3722626038211\\
61.375	8.0683175396467\\
61.5	7.4101718835604\\
61.625	7.3851506100356\\
61.75	7.5649097070799\\
61.875	7.4599416783763\\
62	7.4331880003881\\
62.125	7.5744404240069\\
62.25	7.7974800562133\\
62.375	7.6682336110232\\
62.5	7.7448115962086\\
62.625	7.6716164179944\\
62.75	7.8006080637366\\
62.875	7.7789307342516\\
63	7.5442554415155\\
63.125	7.4752515485596\\
63.25	7.3988815835952\\
63.375	7.4596939925626\\
63.5	7.5652886551834\\
63.625	7.5197889032192\\
63.75	7.5943261754697\\
63.875	7.5304276428701\\
64	7.4929153764224\\
64.125	7.4255828311763\\
64.25	7.4875677779129\\
64.375	7.4234600986958\\
64.5	7.3502446021751\\
64.625	7.3013776895216\\
64.75	7.2664849138512\\
64.875	7.2157012887302\\
65	7.2121595083032\\
65.125	7.2232848240552\\
65.25	7.1877627730417\\
65.375	7.2070707853438\\
65.5	7.2198412091992\\
65.625	7.2727436887236\\
65.75	7.3191280438842\\
65.875	7.4550031216699\\
66	7.4613836445707\\
66.125	7.6249198597028\\
66.25	7.6343647808215\\
66.375	7.6010697361336\\
66.5	7.6554742149002\\
66.625	7.5179723268745\\
66.75	7.818193342965\\
66.875	8.1635896063397\\
67	8.258665708316\\
67.125	8.0265774090215\\
67.25	7.8392090728149\\
67.375	8.1120785247856\\
67.5	8.2686604057035\\
67.625	8.2098250548146\\
67.75	7.9590596694862\\
67.875	8.1083655312182\\
68	8.5335672304054\\
68.125	8.1098373864174\\
68.25	8.3103846918218\\
68.375	8.1503872521908\\
68.5	8.4186239011015\\
68.625	8.507103411132\\
68.75	8.3886728883129\\
68.875	8.0783738564509\\
69	7.9763031287548\\
69.125	8.0316054027258\\
69.25	8.000864620727\\
69.375	8.0462709725026\\
69.5	8.0232935682133\\
69.625	7.9829076781863\\
69.75	7.8795500836728\\
69.875	7.8122563657185\\
70	8.0291463608454\\
70.125	8.0608865668274\\
70.25	8.0187343688358\\
70.375	7.8621057420241\\
70.5	7.7463347410703\\
70.625	7.7518821173569\\
70.75	7.7895685912892\\
70.875	7.8886690724885\\
71	7.7381658134849\\
71.125	7.7989080059738\\
71.25	7.6935824634144\\
71.375	7.5723806322816\\
71.5	7.6097191769058\\
71.625	7.5195773172736\\
71.75	7.5707811540361\\
71.875	7.7116463639614\\
72	7.7789099681432\\
72.125	7.9442730296765\\
72.25	7.5943755505231\\
72.375	7.8063689449423\\
72.5	7.8890952068653\\
72.625	7.6981842748211\\
72.75	7.6734117533104\\
72.875	7.7848345013136\\
73	7.7898073622018\\
73.125	7.7804124783022\\
73.25	7.6316676415881\\
73.375	7.6541293086181\\
73.5	7.6953625100035\\
73.625	7.6802830448294\\
73.75	7.7602009628735\\
73.875	7.7106186107621\\
74	7.721417003766\\
74.125	7.6662077226607\\
74.25	7.5905200636563\\
74.375	7.4255292000578\\
74.5	7.4594763326903\\
74.625	7.2348182329486\\
74.75	7.5528651153281\\
74.875	7.3773163466444\\
75	7.2691884764\\
};
\addlegendentry{Power meter method}

\addplot [color=mycolor3, line width=1.5pt]
  table[row sep=crcr]{%
50	8.73947808459664\\
50.125	8.7370032672046\\
50.25	8.45501801609981\\
50.375	8.53086758873268\\
50.5	8.57230425508892\\
50.625	8.57790561121468\\
50.75	8.81338116097977\\
50.875	8.83762684576343\\
51	8.65025366481427\\
51.125	9.12508487450237\\
51.25	9.61166299607172\\
51.375	9.63093112348474\\
51.5	9.35881887697903\\
51.625	9.24702877889182\\
51.75	8.95150214494438\\
51.875	8.97265056754864\\
52	8.85349137062713\\
52.125	9.31796024278344\\
52.25	8.95994905158594\\
52.375	8.93565613029716\\
52.5	8.99728493829317\\
52.625	9.1502838071872\\
52.75	9.36438799844428\\
52.875	8.66645368204708\\
53	8.62515299241038\\
53.125	8.45515865795157\\
53.25	8.78757374490048\\
53.375	8.72628950927696\\
53.5	8.68157995722899\\
53.625	8.78808059155848\\
53.75	8.76493838099576\\
53.875	8.83946501192515\\
54	8.94833003822819\\
54.125	8.87841757046457\\
54.25	8.80514151527482\\
54.375	8.80866313516923\\
54.5	8.72677601227964\\
54.625	8.81861440289287\\
54.75	8.61283326349758\\
54.875	8.6697424151463\\
55	8.69299588566585\\
55.125	8.81047837847763\\
55.25	9.02398657462382\\
55.375	8.8195648364134\\
55.5	8.60049778142322\\
55.625	8.71365983933692\\
55.75	8.64327666022742\\
55.875	8.59913212423121\\
56	8.67490323901899\\
56.125	8.58269157832667\\
56.25	8.55106569480965\\
56.375	8.46625021749797\\
56.5	8.45039092805665\\
56.625	8.52987649923219\\
56.75	8.60895787168338\\
56.875	8.73733515821136\\
57	8.69713878838482\\
57.125	8.74200251719467\\
57.25	8.84586068274994\\
57.375	8.74429876203735\\
57.5	8.69492532485282\\
57.625	8.61138644172192\\
57.75	8.60976043776724\\
57.875	8.53338144131441\\
58	8.5084883181652\\
58.125	8.31502313497434\\
58.25	8.54660321712038\\
58.375	8.46576224815922\\
58.5	8.53562827247751\\
58.625	8.33028028530008\\
58.75	8.18179786614284\\
58.875	8.5506666452157\\
59	8.25828590683984\\
59.125	8.21396632981638\\
59.25	7.9672726821611\\
59.375	8.19540309549185\\
59.5	7.86037926604329\\
59.625	8.25424675491183\\
59.75	7.78308739071016\\
59.875	8.2969863144132\\
60	8.2497832723577\\
60.125	8.51781065144836\\
60.25	8.43697734038568\\
60.375	8.00669486956218\\
60.5	7.94477558690277\\
60.625	7.97148429508329\\
60.75	8.41413286812143\\
60.875	8.91072048105614\\
61	8.46219285644357\\
61.125	8.36172256853591\\
61.25	7.99762710936774\\
61.375	8.14801513722013\\
61.5	8.15042037885635\\
61.625	8.13731379897736\\
61.75	8.43412957796534\\
61.875	8.11518143660777\\
62	8.24703224698282\\
62.125	8.34542827662978\\
62.25	8.32221782127074\\
62.375	8.30542663762323\\
62.5	8.46327632156786\\
62.625	8.68235949587168\\
62.75	8.42209235107554\\
62.875	8.4444733573194\\
63	8.61953613006483\\
63.125	8.39656292731423\\
63.25	8.65118546973684\\
63.375	8.72932133217843\\
63.5	8.42500851342735\\
63.625	8.32149644947712\\
63.75	8.54128835181941\\
63.875	8.37290076539556\\
64	8.25685214609864\\
64.125	8.56857475821957\\
64.25	8.46578685653772\\
64.375	8.44251869607442\\
64.5	8.384367909336\\
64.625	8.55300928581483\\
64.75	8.46847556672702\\
64.875	8.2110004714385\\
65	8.44490595717488\\
65.125	8.18189940940334\\
65.25	8.38532627300732\\
65.375	8.4076328598662\\
65.5	8.43694310553245\\
65.625	8.43480789210687\\
65.75	8.4820047638947\\
65.875	8.65818896042995\\
66	8.475555774631\\
66.125	8.45269227660591\\
66.25	8.71910376888571\\
66.375	8.34807410641155\\
66.5	8.28681635384394\\
66.625	8.48509358982578\\
66.75	8.89184070999173\\
66.875	8.50901428518467\\
67	8.432865219222\\
67.125	8.5482748552691\\
67.25	8.67743707259192\\
67.375	8.63754606105665\\
67.5	8.87098444764861\\
67.625	8.6883269893128\\
67.75	8.57482034072623\\
67.875	8.5864914922523\\
68	8.45637345141224\\
68.125	8.51740380867106\\
68.25	8.58295539230062\\
68.375	8.55030909486933\\
68.5	8.43901166664307\\
68.625	8.64328318551611\\
68.75	8.90126652279044\\
68.875	8.462320997064\\
69	8.42308027032754\\
69.125	9.00021309383623\\
69.25	8.73146730120884\\
69.375	8.55854824197646\\
69.5	8.9436723598387\\
69.625	8.88950264168538\\
69.75	8.40261956626304\\
69.875	8.40198922831138\\
70	8.38406607341695\\
70.125	8.44895863276853\\
70.25	9.27862720815335\\
70.375	8.64178278209465\\
70.5	8.22604805276008\\
70.625	8.4678129965815\\
70.75	8.45984486744712\\
70.875	8.5680081926153\\
71	8.55060562875651\\
71.125	8.74029271930531\\
71.25	8.34758667397879\\
71.375	8.43099396716297\\
71.5	8.24466569515184\\
71.625	8.15566605515289\\
71.75	8.22579866883453\\
71.875	8.75530851140686\\
72	8.33660335926738\\
72.125	8.03972846518627\\
72.25	8.11398883760559\\
72.375	8.07371077540559\\
72.5	8.00762939941956\\
72.625	7.97977029579436\\
72.75	8.01030110320431\\
72.875	8.09319843425353\\
73	7.97039771287353\\
73.125	8.06174097326875\\
73.25	8.02051293054562\\
73.375	8.1142569120737\\
73.5	8.24555350474569\\
73.625	8.27950712635131\\
73.75	8.24053568476048\\
73.875	8.43569833655589\\
74	8.42191357198013\\
74.125	8.44261708224929\\
74.25	8.33321722262692\\
74.375	8.23903471615375\\
74.5	8.15711354812624\\
74.625	8.10331654336039\\
74.75	8.18151223705622\\
74.875	8.25197545536509\\
75	8.35660274986343\\
};
\addlegendentry{3-mixer method}

\end{axis}

\node [above right] at (rel axis cs:0.3,1.15) {\includegraphics[width=3cm]{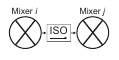}};
\end{tikzpicture}%

%% file: main.bbl
\begin{thebibliography}{10}
\providecommand{\url}[1]{#1}
\csname url@samestyle\endcsname
\providecommand{\newblock}{\relax}
\providecommand{\bibinfo}[2]{#2}
\providecommand{\BIBentrySTDinterwordspacing}{\spaceskip=0pt\relax}
\providecommand{\BIBentryALTinterwordstretchfactor}{4}
\providecommand{\BIBentryALTinterwordspacing}{\spaceskip=\fontdimen2\font plus
\BIBentryALTinterwordstretchfactor\fontdimen3\font minus \fontdimen4\font\relax}
\providecommand{\BIBforeignlanguage}[2]{{%
\expandafter\ifx\csname l@#1\endcsname\relax
\typeout{** WARNING: IEEEtran.bst: No hyphenation pattern has been}%
\typeout{** loaded for the language `#1'. Using the pattern for}%
\typeout{** the default language instead.}%
\else
\language=\csname l@#1\endcsname
\fi
#2}}
\providecommand{\BIBdecl}{\relax}
\BIBdecl

\bibitem{Herold1942}
E.~Herold, ``The operation of frequency converters and mixers for superheterodyne reception,'' \emph{Proceedings of the {IRE}}, vol.~30, no.~2, pp. 84--103, Feb. 1942, {DOI}: \href{http://dx.doi.org/10.1109/jrproc.1942.233661}{10.1109/jrproc.1942.233661}.

\bibitem{hayton2013}
D.~J. Hayton, A.~Khudchencko, D.~G. Pavelyev, J.~N. Hovenier, A.~Baryshev, J.~R. Gao, T.~Y. Kao, Q.~Hu, J.~L. Reno, and V.~Vaks, ``Phase locking of a 3.4 {THz} third-order distributed feedback quantum cascade laser using a room-temperature superlattice harmonic mixer,'' \emph{Appl. Phys. Lett.}, vol. 103, Aug. 2013, {Art.} no. 051115, {DOI}:~\href{https://doi.org/10.1063/1.4817319}{10.1063/1.4817319}.

\bibitem{Curwen2023}
C.~A. Curwen, J.~H. Kawamura, D.~J. Hayton, S.~J. Addamane, J.~L. Reno, B.~S. Williams, and B.~S. Karasik, ``{Phase Locking of a THz QC-VECSEL to a Microwave Reference},'' \emph{IEEE Transactions on Terahertz Science and Technology}, vol.~13, no.~5, pp. 448--453, 2023, {DOI}: \href{http://dx.doi.org/10.1109/TTHZ.2023.3280451}{10.1109/TTHZ.2023.3280451}.

\bibitem{maas}
S.~Maas, \emph{Microwave mixers. Artech House, Inc}.\hskip 1em plus 0.5em minus 0.4em\relax Artech House Inc,, 1986.

\bibitem{Trambarulo}
R.~Trambarulo and H.~Berger, ``{Conversion Loss and Noise Temperature of Mixers from Noise Measurements},'' in \emph{1983 IEEE MTT-S International Microwave Symposium Digest}, 1983, pp. 364--365, {DOI}: \href{http://dx.doi.org/10.1109/MWSYM.1983.1130913}{10.1109/MWSYM.1983.1130913}.

\bibitem{Dunsmore}
J.~Dunsmore, ``Novel method for vector mixer characterization and mixer test system vector error correction,'' in \emph{2002 IEEE MTT-S International Microwave Symposium Digest (Cat. No.02CH37278)}, vol.~3, 2002, pp. 1833--1836 vol.3, {DOI}: \href{http://dx.doi.org/10.1109/MWSYM.2002.1012219}{10.1109/MWSYM.2002.1012219}.

\bibitem{Alireza2013}
A.~Kazemipour, M.~Salhi, T.~Kleine-Ostmann, and T.~Schrader, ``{Novel method to measure the conversion-losses (C.L.) of microwave and mm-wave mixers},'' in \emph{2013 Asia-Pacific Microwave Conference Proceedings (APMC)}, 2013, pp. 731--733, {DOI}: \href{http://dx.doi.org/10.1109/APMC.2013.6694912}{10.1109/APMC.2013.6694912}.

\bibitem{bulcha2016}
B.~T. {Bulcha}, J.~L. {Hesler}, V.~{Drakinskiy}, J.~{Stake}, A.~{Valavanis}, P.~{Dean}, L.~H. {Li}, and N.~S. {Barker}, ``\textcolor{black}{Design and Characterization of 1.8–3.2 {THz} {Schottky}-Based Harmonic Mixers},'' \emph{IEEE Trans. THz Sci. Technol.}, vol.~6, no.~5, pp. 737--746, Sep. 2016, {DOI}:~\href{https://doi.org/10.1109/TTHZ.2016.2576686}{10.1109/TTHZ.2016.2576686}.

\bibitem{divya2021}
D.~Jayasankar, V.~Drakinskiy, N.~Rothbart, H.~Richter, X.~Lu, L.~Schrottke, H.~T. Grahn, M.~Wienold, H.-W. Hübers, P.~Sobis, and J.~Stake, ``{A 3.5-THz, $\times$6-Harmonic, Single-Ended Schottky Diode Mixer for Frequency Stabilization of Quantum-Cascade Lasers},'' \emph{IEEE Transactions on Terahertz Science and Technology}, vol.~11, no.~6, pp. 684--694, 2021, {DOI}:~\href{https://doi.org/10.1109/TTHZ.2021.3115730}{10.1109/TTHZ.2021.3115730}.

\bibitem{Reck2023}
T.~Reck, S.~Durant, and J.~Hesler, ``{Design of a 2.5 THz Schottky-Diode Fourth-Harmonic Mixer},'' \emph{IEEE Transactions on Terahertz Science and Technology}, vol.~13, no.~6, pp. 580--586, 2023, {DOI}: \href{http://dx.doi.org/10.1109/TTHZ.2023.3307566}{10.1109/TTHZ.2023.3307566}.

\bibitem{Newell73}
A.~Newell, R.~Baird, and P.~Wacker, ``Accurate measurement of antenna gain and polarization at reduced distances by an extrapolation technique,'' \emph{IEEE Transactions on Antennas and Propagation}, vol.~21, no.~4, pp. 418--431, 1973, {DOI}:~\href{https://doi.org/10.1109/TAP.1973.1140519}{10.1109/TAP.1973.1140519}.

\bibitem{Beatty67}
R.~Beatty, ``Discussion of errors in gain measurements of standard electromagnetic horns,'' \emph{NBS, Boulder, Colo., Tech. Note 351}, 1967.

\bibitem{waveguide}
``{IEEE Standard for Rectangular Metallic Waveguides and Their Interfaces for Frequencies of {110 GHz} and Above--Part 1: Frequency Bands and Waveguide Dimensions},'' \emph{{IEEE} Std 1785.1-2012}, pp. 1--22, Mar. 2013, {DOI}:~\href{https://doi.org/10.1109/IEEESTD.2013.6471987}{10.1109/IEEESTD.2013.6471987}.

\bibitem{Li2014}
H.~Li, A.~Arsenovic, J.~L. Hesler, A.~R. Kerr, and R.~M. Weikle, ``\textcolor{black}{ Repeatability and Mismatch of Waveguide Flanges in the 500–750 GHz Band},'' \emph{IEEE Transactions on Terahertz Science and Technology}, vol.~4, no.~1, p. 39–48, Jan. 2014, {DOI}:~\href{https://doi.org/10.1109/tthz.2013.2283540}{10.1109/tthz.2013.2283540}.

\bibitem{goldsmith}
P.~F. {Goldsmith}, ``{{Quasi-optical techniques}},'' \emph{Proc. IEEE}, vol.~80, no.~11, pp. 1729--1747, Nov. 1992, {DOI}:~\href{https://doi.org/10.1109/5.175252}{10.1109/5.175252}.

\bibitem{Vanmoer2007}
W.~Van~Moer and Y.~Rolain, ``Determining the reciprocity of mixers through three-port large signal network analyzer measurements,'' \emph{IEEE Transactions on Instrumentation and Measurement}, vol.~56, no.~5, pp. 2051--2056, 2007, {DOI}: \href{http://dx.doi.org/10.1109/TIM.2007.903645}{10.1109/TIM.2007.903645}.

\bibitem{Fujii2012}
K.~Fujii, T.~Tosaka, K.~Fukunaga, and Y.~Matsumoto, ``{RF} power measurement in {D}-band using down-converter calibrated by three-mixer method,'' \emph{IEICE Electronics Express}, vol.~9, no.~13, pp. 1096--1101, 2012, {DOI}: \href{http://dx.doi.org/10.1587/elex.9.1096}{10.1587/elex.9.1096}.

\bibitem{Tosaka2012}
T.~Tosaka, K.~Fujii, K.~Fukunaga, and Y.~Matsumoto, ``{Measurement of frequency conversion losses with 3-mixer method for traceable mm-wave power measurement method in D-band},'' in \emph{2012 37th International Conference on Infrared, Millimeter, and Terahertz Waves}, 2012, pp. 1--2, {DOI}: \href{http://dx.doi.org/10.1109/IRMMW-THz.2012.6380427}{10.1109/IRMMW-THz.2012.6380427}.

\bibitem{Johansson_1992}
J.~Johansson and N.~Whyborn, ``The diagonal horn as a sub-millimeter wave antenna,'' \emph{IEEE Transactions on Microwave Theory and Techniques}, vol.~40, no.~5, pp. 795--800, 1992, {DOI}: \href{http://dx.doi.org/10.1109/22.137380}{10.1109/22.137380}.

\bibitem{VDI}
\BIBentryALTinterwordspacing
VDI. {Integrated AMC - Virginia Diodes Inc. model: WR1.5AMC-I}. [Online]. Available: \url{https://www.vadiodes.com/index.php/en/products/compact-transmitter-modules-amc-i?id=922}
\BIBentrySTDinterwordspacing

\bibitem{Penfield}
R.~Bauer and P.~Penfield, ``De-embedding and unterminating,'' \emph{IEEE Transactions on Microwave Theory and Techniques}, vol.~22, no.~3, pp. 282--288, 1974, {DOI}: \href{http://dx.doi.org/10.1109/TMTT.1974.1128212}{10.1109/TMTT.1974.1128212}.

\bibitem{Savitzky1964}
A.~Savitzky and M.~J.~E. Golay, ``Smoothing and differentiation of data by simplified least squares procedures,'' \emph{Analytical Chemistry}, vol.~36, no.~8, pp. 1627--1639, Jul. 1964, {DOI}: \href{http://dx.doi.org/10.1021/ac60214a047}{10.1021/ac60214a047}.

\bibitem{Morgan2013}
M.~A. Morgan and S.-K. Pan, ``{\textcolor{black}{Graphical Prediction of Trapped Mode Resonances in Sub-mm and THz Waveguide Networks}},'' \emph{{IEEE Transactions on Terahertz Science and Technology}}, vol.~3, no.~1, pp. 72--80, 2013, {DOI}:~\href{https://doi.org/10.1109/TTHZ.2012.2235910}{10.1109/TTHZ.2012.2235910}.

\bibitem{Friis1946}
H.~Friis, ``\textcolor{black}{A Note on a Simple Transmission Formula},'' \emph{Proceedings of the IRE}, vol.~34, no.~5, p. 254–256, May 1946, {DOI}:~\href{https://doi.org/10.1109/jrproc.1946.234568}{10.1109/jrproc.1946.234568}.

\bibitem{VDI2}
\BIBentryALTinterwordspacing
VDI. {Sub-harmonic mixer - Virginia Diodes Inc. model: WR15SHM}. [Online]. Available: \url{https://www.vadiodes.com/en/products/mixers-shm-ehm-and-fm?id=400}
\BIBentrySTDinterwordspacing

\end{thebibliography}
